\documentclass[a4paper,12pt]{article}
\usepackage[left=2.5cm,bottom=3cm,right=2.5cm,top=3cm]{geometry}

\usepackage{overpic,youngtab}
\usepackage{subfigure}
\usepackage[latin,english]{babel}
\usepackage{amsmath}
\usepackage{amssymb}
\usepackage{epstopdf}
\usepackage{graphics,psfrag,rotating}
\usepackage{mathabx}
\usepackage{graphicx}
\usepackage{dcolumn}
\usepackage{float}
\usepackage{pdflscape}
\usepackage{cancel}
\usepackage{array}
\usepackage{booktabs}
\usepackage{amscd}
\usepackage{mathtools}
\usepackage{fancybox}
\usepackage{tensor}
\usepackage{fix-cm}
\usepackage[colorlinks=true,citecolor=blue,,linktocpage=true,linkcolor=blue,urlcolor=black]{hyperref}
\usepackage{cleveref}
\usepackage{tikz}
\usetikzlibrary{patterns,snakes}
\tikzstyle{spring}=[line width=0.8,blue!7!black!80,snake=coil,segment amplitude=4.25,segment length=4.75,line cap=round]
\usetikzlibrary{decorations.pathmorphing}
\usetikzlibrary{matrix}
\usetikzlibrary{positioning}
\usetikzlibrary{arrows}
\usepackage{titlesec}
\usepackage{abstract}
\usepackage{cancel}
\usepackage{mathtools}
\usepackage{pdflscape}
\usepackage{graphicx}

\def\be{\begin{equation}}
\def\ee{\end{equation}}
\def\bea{\begin{eqnarray}}
\def\eea{\end{eqnarray}}
\def\pd{\partial}
\def\a{\alpha}
\def\b{\beta}
\def\g{\gamma}
\def\d{\delta}
\def\m{\mu}
\def\n{\nu}

\def\l{\lambda}

\def\r{\rho}

\def\s{\sigma}
\def\e{\epsilon}
\def\bi{\begin{itemize}}
	\def\ei{\end{itemize}}
\def\bg{\bar{g}}

\begin{document}
	\vspace*{-1cm}
\phantom{hep-ph/***}
{\flushleft
	{{FTUAM-21-xx}}
	\hfill{{ IFT-UAM/CSIC-21-96}}}
\vskip 1.5cm
\begin{center}
	{\LARGE\bfseries The quantum dynamics of Lagrange multipliers.}\\[3mm]
	\vskip .3cm
	
\end{center}

\vskip 0.5  cm
\begin{center}
	{\large Enrique \'Alvarez$^{\dagger}$, Jes\'us Anero$^\dagger$, Carmelo P. Martin$^{\dagger\dagger}$ and Eduardo Velasco-Aja$^\dagger$.}
	\\
	\vskip .7cm
	{
		$\dagger$Departamento de F\'isica Te\'orica and Instituto de F\'{\i}sica Te\'orica,
		IFT-UAM/CSIC,\\
		Universidad Aut\'onoma de Madrid, Cantoblanco, 28049, Madrid, Spain\\
        $\dagger\dagger$Universidad Complutense de Madrid (UCM), Departamento de F\'isica Te\'orica and \\
        IPARCOS, Facultad de Ciencias F\'isicas, 28040 Madrid, Spain
		\vskip .1cm

		\vskip .5cm
		
		\begin{minipage}[l]{.9\textwidth}
			\begin{center}
				\textit{E-mail:}
				\tt{enrique.alvarez@uam.es},
				\tt{jesusanero@gmail.com},
                \tt{carmelop@fis.ucm.es} and
				\tt{eduardo.velasco@uam.es}
			\end{center}
		\end{minipage}
	}
\end{center}
\thispagestyle{empty}

\begin{abstract}
	\noindent
When implementing a non-linear constraint in quantum field theory by means of a Lagrange multiplier, $\l(x)$, it is often the case that quantum dynamics induce quadratic and even higher order terms in $\l(x)$, which then does not enforce the constraint anymore.
This is illustrated in the case of Unimodular Gravity, where the constraint is that the metric tensor has to be unimodular ($g(x)\equiv \det\,g_{\m\n}(x)=-1$).
\end{abstract}

\newpage
\tableofcontents
\thispagestyle{empty}
\flushbottom

\newpage

\section{Introduction{.}}
Unimodular gravity (UG) is a variant of General Relativity (GR) originally conceived by Einstein in 1919 (cf. \cite{Alvarez} for a recent review) in which the zero mode component of the vacuum energy does not weigh.
This is the only fully satisfactory solution to at least part of the well-known Cosmological Constant (CC) problem.
\par
The theory rests on the assumption that only unimodular metrics ($ g{(x)}\equiv\det\,g_{\m\n}{(x)}=-1$) are admissible.
This poses a formidable problem in  practice because it is a non-linear constraint. There have been at least three attempts to implement this constraint, namely:
\begin{enumerate}
\item The simplest one consists in supplementing the GR lagrangian with a Lagrange multiplier term implementing the constraint {either as}
\begin{equation}
\Delta S_{UG}\equiv \int d^d x \,{1\over \kappa}\l(x)\left(g{(x)}+1\right)\quad\text{or {as}}\quad  {\Delta S_{UG}}\equiv\int d^d x \,{1\over \kappa}\l(x)\left( \sqrt{{-}g{(x)}}-1\right),
\label{firstapp}
\end{equation}
along with the standard linear splitting $g_{\mu\nu}{(x)}=\bar{g}_{\mu\nu}{(x)}+\kappa\,h_{\mu\nu}{(x)}$.
 A recent example of this is to be found in \cite{Kugo}. Note that the mass dimension of the multiplier is $[\l(x)]=3$ in $d=4$.
\item The second one just defines an auxiliary unrestricted metric $g_{\m\n}$ in terms of which the unimodular metric is obtained
\begin{equation}
\g_{\m\n}{(x)}\equiv g^{- 1/d}{(x)}\,g_{\m\n}{(x)}.
\label{Enriques}
\end{equation}
This introduces a new Weyl gauge symmetry under
\begin{equation}
g_{\m\n}{(x)}\rightarrow \Omega^2(x) g_{\m\n}{(x)}.
\end{equation}
This formalism has been used extensively in \cite{Alvarez:2015sba}.
\item Finally, use can be made of the theorem on the effect that any unimodular metric is the exponential of a traceless one
\begin{equation}
g_{\m\n}{(x)}=e^{G_{\m\n}(x)},
\end{equation}
with
\begin{equation}
\text{tr}\,G_{\m\n}(x)=0.
\end{equation}
This has been used in \cite{deLeonArdon, Percacci, Eichhorn, deBrito}.
\end{enumerate}
The purpose of the present work is to examine the consistency of the first alternative under quantum corrections, although some comments will also be made on the exponential parametrization.
\par
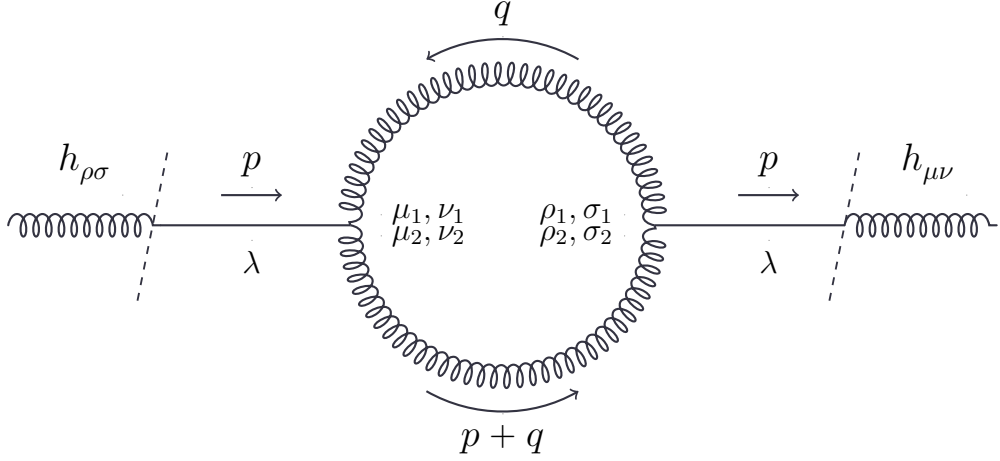
\begin{figure}[h!]
  \centering
  \begin{tikzpicture}[scale=2,
   decoration={coil,amplitude=4.25,segment length=4.75}]
   \filldraw [black] (-2.65,0.25) circle (0pt) node[anchor=south]{\large$p$};
   \draw[thick, ->,blue!7!black!80](-2.85,0.2) -- (-2.45,0.2);
   \filldraw [black] (0.75,0.25) circle (0pt) node[anchor=south]{\large$p$};
   \filldraw [black] (-2.65,-0.1) circle (0pt) node[anchor=north]{$\lambda$};
   \filldraw [black] (0.75,-0.1) circle (0pt) node[anchor=north]{$\lambda$};
   \draw[thick, <-,blue!7!black!80](0.95,0.2) -- (0.55,0.2);
   \draw[thick, ->,blue!7!black!80](-0.5,1.1) arc (60:120:1);
   \draw[thick, <-,blue!7!black!80](-0.5,-1.1) arc (-60:-120:1);
   \draw[spring] (-4.25,0) -- (-3.25,0);
   \draw[line width=0.8,blue!7!black!80] (-3.25,0) -- (-2,0);
   \draw[decorate,line width=0.8,blue!7!black!80] (0,0) arc (0:180:1);
   \draw[decorate,line width=0.8,blue!7!black!80] (-2,0) arc (-180:0:1);
   \draw[line width=0.8,blue!7!black!80] (0,0) -- (1.25,0);
   \draw[spring] (1.25,0) -- (2.25,0);
   \filldraw [black] (-1,1.25) circle (0pt) node[anchor=south]{\large$q$};
   \filldraw [black] (-1.8,0.07) circle (0pt) node[anchor=west]{$\mu_1,\nu_1$};
   \filldraw [black] (-1.8,-0.07) circle (0pt) node[anchor=west]{$\mu_2,\nu_2$};
   \filldraw [black] (-0.2,0.07) circle (0pt) node[anchor=east]{$\rho_1,\sigma_1$};
   \filldraw [black] (-0.2,-0.07) circle (0pt) node[anchor=east]{$\rho_2,\sigma_2$};
   \filldraw [black] (-1,-1.25) circle (0pt) node[anchor=north]{\large$p+q$};
   \draw[line width=0.7,blue!7!black!80,dashed] (-3.4,-0.5) -- (-3.2,0.5);
   \draw[line width=0.7,blue!7!black!80,dashed] (1.15,-0.5) -- (1.35,0.5);
   \filldraw [black] (-3.5,0.2) circle (0pt) node[anchor=south east]{\large$\tensor{h}{_\rho_\sigma} $};
   \filldraw [black] (2.05,0.2) circle (0pt) node[anchor=south east]{\large$\tensor{h}{_\mu_\nu} $};

  \end{tikzpicture}
   \caption{Feynman diagram yielding a divergent $\lambda^2(x  )$ part.}\label{fig:1}
   \end{figure}

It is quite intuitive that the coupling of the multiplier with the graviton can induce divergent diagrams of higher order in the {multipliers} and even{,} in some cases kinetic energy terms for them{; see \cref{fig:1,fig:2} below}. Examples of related phenomena appear in the principal chiral model \cite{Polyakov} and in the physics of gravitons in a codimension one brane \cite{Dvali}, among others.\par
The conclusion is that any consistent renormalization\footnote{Somebody could be tempted to put by hand all these coupling constants to zero. This would {be} most unnatural; it is more  {or} less equivalent to {putting} to zero all coupling constants in front of the higher dimensional operators in quantum gravity{; no known symmetry principle supports this.}}
must include finite values for the coupling constants in front of these operators, which in turn conveys the fact that the Lagrange multiplier does not work as a multiplier anymore: it has become a full fledge field with its own dynamics.
\par
We {shall,} in fact, demonstrate in detail how the operator
\begin{equation}
 {\cal O}_1\equiv \l(x)^2,
\end{equation}
is generated by one-loop diagrams. { We further argue, without explicit calculation,} that a two-loop diagram will induce the kinetic energy correction
\begin{equation}
{\cal O}_2\equiv \left(\pd_\m\l(x)\right)^2.
\end{equation}


To be specific, let $g {(x)}$ denote{,} as usual{,} the determinant of the metric. Assume that the unimodularity condition; $g{(x)}=-1$, is enforced with the first alternative just exposed, {i.e.} by including in the path integral the additional contribution,
\begin{equation}
  \int {\cal D} \lambda\; e^{\frac{i}{\kappa}\int_{}^{}d^{4}x   \; \lambda(x)\left(g{(x)}+1\right) }. \label{mult}
\end{equation}
We want to explore the implications of such a \textit{gauge fixing} procedure when radiative corrections are taken into  {account} in the covariant perturbation theory of quantum gravity.
Consider gravitons, $\tensor{h  }{_\mu_\nu} (x) $ propagating on a flat background, $\tensor{\eta}{_\mu_\nu} $, such that,
\begin{equation}
  \tensor{g  }{_\mu_\nu}(x) =\tensor{\eta}{_\mu_\nu}+ \kappa \,\tensor{h  }{_\mu_\nu} (x).
\end{equation}
Since,
\begin{align}
  -g(x)&= 1+ \kappa \,h(x)+ \frac{\kappa^2}{2}\left(h^2(x)-\tensor{h  }{_\mu^\nu} (x)\tensor{h  }{^\mu_\nu} (x)\right) + {O}\left(\kappa\right)^3,
\end{align}
where $h(x)=\tensor{h}{_\mu_\nu}(x)\tensor{\eta}{^\mu^\nu}$, the term in \cref{mult} gives rise to an interaction
\begin{align}
  &i \frac{\kappa}{2}\int_{}^{}d^{4} x \; \lambda(x)\tensor{h}{_{\mu_1}_{\nu_1}}(x)\tensor{h}{_{\mu_2}_{\nu_2}}(x)\tensor{V}{_L^{\mu_1}^{\nu_1}^{\mu_2}^{\nu_2}},\\
  &\tensor{V}{_L^{\mu_1}^{\nu_1}^{\mu_2}^{\nu_2}}=\frac{1}{2}\left(2 \tensor{\eta}{^{\mu_1}^{\nu_1}} \tensor{\eta}{^{\mu_2}^{\nu_2}}-\tensor{\eta}{^{\mu_1}^{\mu_2}} \tensor{\eta}{^{\nu_1}^{\nu_2}}-\tensor{\eta}{^{\mu_1}^{\mu_2}} \tensor{\eta}{^{\nu_1}^{\nu_2}}\right) .\label{vert}
\end{align}
At  {the} tree-level, the contribution to the 1PI function is linear in $\lambda{(x)}$ for it is given by \cref{mult}. However, radiative corrections yield{,} in fact{,} a one-loop contribution to the 1PI functional quadratic in $\lambda{(x)}$.
Such contribution is given by the 1PI diagram in \cref{fig:1}, whose value in dimensional regularization is given by the following  {F}eynman integral
\begin{equation}
  \Gamma_L(p)=\frac{\kappa^2}{2}\int_{}^{}\frac{d^{d } q}{(2 \pi)^d}\tensor{V}{_L^{\mu_1}^{\nu_1}^{\mu_2}^{\nu_2}}\tensor{G}{_{\mu_1}_{\nu_1}_{\rho1}_{\sigma1}}(p) \tensor{G}{_{\mu_2}_{\nu_2}_{\rho2}_{\sigma2}}(p+q) \tensor{V}{_L^{\rho1}^{\sigma1}^{\rho2}^{\sigma2}},\label{diag}
\end{equation}
where the vertices $V_L $ are given by \cref{vert} and $\tensor{G}{_{\mu_1}_{\nu_1}_{\rho1}_{\sigma1}}(p)$ denotes the graviton propagator discussed in \cref{sec:2}.

\section{Computing the diagram.}\label{sec:2}
In this section we shall work out $\Gamma_L(p)$ in \cref{diag} for quantum unimodular gravity as defined in \cite{Alvarez:2015sba} on the one side and \cite{Kugo} on the other.

The general structure of the propagator is of the form,
\begin{align}
&G_{\mu_1\nu_1\mu_2\nu_2}(q)=-\frac{1}{q^2}\Bigg(A_1\,(\eta_{\mu_1\mu_2}\eta_{\nu_1\nu_2}+\eta_{\mu_1\nu_2}\eta_{\nu_1\mu_2})+A_2\eta_{\mu_1\nu_1}\eta_{\mu_2\nu_2}
+A_3\,\frac{1}{q^2}\Big(\eta_{\mu_1\nu_1}q_{\mu_2}q_{\nu_2}+\nonumber\\
+&\eta_{\mu_2\nu_2}q_{\mu_1}q_{\nu_1}\Big)+A_4\,\frac{1}{q^2}(\eta_{\mu_1\mu_2}q_{
  \nu_1}q_{\nu_2}+\eta_{\mu_1\nu_2}q_{\nu_1}q_{\mu_2}+\eta_{\nu_1\mu_2}q_{\mu_1}q_{\nu_2}+\eta_{\nu_1\nu_2}q_{\mu_1}q_{\mu_2})+\nonumber\\
  +&A_5\,
  \frac{1}{q^4}q_{\mu_1}q_{\nu_1}q_{\mu_2}q_{\nu_2}\Bigg).
  \label{gravitonprop}
  \end{align}
The value of the coefficients $A_1$ to $A_5$ depends, in general, on the set of fields introduced by the BRST quantization procedure. The set of fields introduced in \cite{Alvarez:2015sba} is quite different from the set of fields in \cite{Kugo}. Let us then begin by adapting the BRST formalism of \cite{Alvarez:2015sba}, which was developed for the unrestricted metric in \cref{Enriques}, to our case.


 The action of the quantum theory is not invariant under the full diffeomorphism group but only under the TDiff subgroup{;}.
\begin{equation}\d g_{\m\n}=\nabla_\m c^{T}_\n+\nabla_\n c^{T}_\m \hspace{1cm}\nabla_\l c^{T\l}=0.\end{equation}
The nilpotent BRST operator $s^2_D=0$ is defined by
\begin{align}
&s_D g_{\m\n}=0,\nonumber\\
&s_D h_{\m\n}=\partial_\m c^T_\n+\partial_\n c^T_\m+c^{T\r}\partial_\r h_{\m\n}+h_{\r\n}\partial_\m c^{T\r}+h_{\r\m}\partial_\n c^{T\r},\end{align}
and
\begin{align}
&s_D g{(x)}=2g{(x)}\nabla_\m c^{T\m}=c^{T\m}\partial_\m g{(x)},\nonumber\\
&s_D\l{(x)}=c^{T\r}\partial_\r \l{(x)},
\end{align}
where $c^{T\m}$ is the ghost fields for transverse diffeomorphisms, the transverse condition implies $\partial_\m c^{T\m}=0$. The BRST transformations are then defined in such a way that the BRST algebra closes.

First of all, let us note that
\begin{align}\int d^dx s^2_D\l{(x)}&=-\int d^d x\partial_\r\left({c}^{T\r}{c}^{T\s}\partial_\s\l{(x)}\right)=0.\end{align}

The action is invariant under $s_D$ because
\begin{equation}s_D\int d^d x \l{(x)}\left(g{(x)}+1\right)=\int d^d x\partial_\r\Big[c^{T\r}(g{(x)}+1)\l{(x)}\Big].\end{equation}

Introduce now the following set of fields
\begin{align}\label{f}
&h_{\m\n}^{(0,0)},\,c_\m^{(1,1)},\,b_\m^{(1,-1)},\,f_\m^{(0,0)},\,\phi^{(0,2)},\nonumber\\
&\pi^{(1,-1)},\,\pi'^{(1,1)},\,\bar{c}^{(0,-2)},\,c'^{(0,0)},\nonumber\\
&c^{(1,1)},\,b^{(1,-1)},\,f^{(0,0)},
\end{align}
where $c_\m^{(1,1)}$ denotes $c_\m$ and $h_{\m\n}^{(0,0)}$ is $h_{\m\n}$ and the superscript $(n,m)$ carries the Grassmann number, $n$ (defined modulo two) and ghost number, $m$. Also we need to introduce a projector $\Theta_{\m\n}$
\begin{equation}\label{CT} c_\m^T=\Theta_{\m\n}c^\n=\left(\eta_{\m\n}\Box-\partial_\m\partial_\n\right)c^{\n (1,1)},\end{equation}
and
\begin{equation}\partial_\m\left(c^{\r T}\partial_\r c^{T\m}\right)=0,\, \partial_\m\left[(Q^{-1})^\m_\n\left(c^{\r T}\partial_\r c^{T\m}\right)\right]=0,\end{equation}
where
\begin{equation}(Q^{-1})^\m_\n=\frac{1}{\Box}\d^\m_\n.\end{equation}
To summarize, the action of $s_D$ over the fields \eqref{f}
\begin{align}
s_D g_{\m\n}&=0,\nonumber\\
s_D h_{\m\n}&=\partial_\m c^T_\n+\partial_\n c^T_\m+c^{\r T}\partial_\r h_{\m\n}+h_{\r\n}\partial_\m c^{T\r}+h_{\r\m}\partial_\n c^{T\r},\nonumber\\
s_Dc^{(1,1)\m}&=(Q^{-1})^\m_\n\left(c^{\r T}\partial_\r c^{T\n}\right)+\partial^\m \phi^{(0,2)},\nonumber\\
s_D\phi^{(0,2)}&=0,\nonumber\\
s_Db_\m^{(1,-1)}&=f_\m^{(0,0)},\nonumber\\
s_Df_\m^{(0,0)}&=0,\nonumber\\
s_D\bar{c}^{(0,-2)}&=\pi^{(1,-1)},\nonumber\\
s_D\pi^{(1,-1)}&=0,\nonumber\\
s_Dc'^{(0,0)}&=\pi'^{(1,1)},\nonumber\\
s_D\pi'^{(1,1)}&=0,\nonumber\\
s_Dc^{(1,1)}&=c^{T\r}\partial_\r c^{(1,1)},\nonumber\\
s_Db^{(1,-1)}&=c^{T\r}\partial_\r b^{(1,-1)},\nonumber\\
s_Df^{(0,0)}&=c^{T\r}\partial_\r f^{(0,0)}.\label{tab}
\end{align}

The fermion $X_{TD}$ performing the gauge fixing of the TDiff symmetry reads
\begin{equation}X_{TD}=b_\m^{(1,-1)}\left[F^\m+\r_1f^{\m(0,0)}\right]+\bar{c}^{(0,-2)}\left[F_2^\m c_\m^{(1,1)}+\r_2\pi'^{(1,1)}\right]+c'^{(0,0)}\left[F_1^\m
b_\m^{(1,-1)}+\r_3\pi^{(1,-1)}\right],\end{equation}
where $F^\m$ is a function containing the graviton field and $F_1^\m$, $F_2^\m$ and the three operators $\r_i$ can be freely chosen.

For the computation at hand{,} one can choose,
\begin{align}\label{F}&F_\m=\g_1\partial^\n h_{\m\n}+\g_2\partial_\m h,\nonumber\\
&F_1^\m=\a_1\partial^\m,\nonumber\\
&F_2^\m=\a_2\partial^\m,\nonumber\\
&(\r_2-\r_3)^{-1}=\g\Box.\end{align}

Applying the $s_D$ operator over the gauge fixing term using \cref{tab}
\begin{align}\int d^dx\,s_D X_{TD}&=\int d^dx\Bigg\{f_\m^{(0,0)}\left[F^\m+\r_1f^{\m(0,0)}\right]-b_\m^{(1,-1)}s_DF^\m+\pi^{(1,-1)}\left[F_2^\m c_\m^{(1,1)}+\r_2\pi'^{(1,1)}\right]+\nonumber\\
&+\bar{c}^{(0,-2)}F_2^\m\partial_\m\phi^{(0,2)}+\pi'^{(1,1)}\left[F_1^\m
b_\m^{(1,-1)}+\r_3\pi^{(1,-1)}\right]+c'^{(0,0)}F_1^\m f_\m^{(0,0)}\Bigg\},\end{align}
in which,
\begin{align} s_D F^\m&=\g_1\Box c^{T\m}+\g_1\partial_\n\Big[c^{\r T}\partial_\r h^{\m\n}+h^{\r\n}\partial^\m c^{T\r}+h^{\r\m}\partial^\n c^{T\r}\Big]+\nonumber\\
&+\g_2\partial^\m\Big[c^{T\r}\partial_\r h+2h_{\r\s}\partial^\r c^{T\s}\Big].\end{align}
The bosonic quadratic action in our problem reads,
\begin{align} &S_{\text{\tiny{2}}}-S^{TD}_{\text{\tiny{hf}}}= -\frac{1}{2}\int
d^d x\Bigg\{h^{\a\b}\Big[\frac{1}{4}\eta_{\b\n}\eta_{\a\m}\bar{\Box}-\frac{1}{4}\eta_{\a\b}\eta_{\m\n}\bar{\Box}+\frac{1}{2}\eta_{\a\b}\partial_\m\partial_\n-\frac{1}{2}\eta_{\b\n}\partial_\a\partial_\m\Big]h^{\m\n}-2h\l-\nonumber\\
&-2h^{\m\n}\Big[\frac{\g_1}{2}\left(\partial_\n f_\m^{(0,0)}+\partial_\m f_\n^{(0,0)}\right) +\g_2\eta_{\m\n}\partial^\l f_\l^{(0,0)}\Big]+2\r_1f_\m^{(0,0)}f^{\m(0,0)}+2\a_1c'^{(0,0)}\partial^\m f_\m^{(0,0)}\Bigg\},
\end{align}
which has the general structure
\begin{align}
&S_{\text{\tiny{2}}}-S^{TD}_{\text{\tiny{hf}}}=-\frac{1}{2}\,\int\,d^dx\,\psi^A\,K_{AB}\psi^B,
\end{align}
where we have written the quadratic operator corresponding to the generalized field, $\psi^A$, defined as a vector
\begin{equation}
\psi^A\equiv\begin{pmatrix}
	h_{\m\n}\\f^{(0,0)}_\m\\c'^{(0,0)}\\\l
\end{pmatrix}.
\end{equation}
After going into momentum space, the graviton propagator can be obtained by finding the inverse of the $K_{AB}(p)$ operator,
\begin{equation}
  \label{KG}K_{AB}G^{BC}=I_A^C.
\end{equation}
By doing so, one finds that
\begin{align}
  &G_{\r\s\m\n}{(p)}={-\frac{1}{p^2}\Bigg(2\left(\eta_{\r\m}\eta_{\s\n}+\eta_{\r\n}\eta_{\s\m}\right)-\frac{4}{(d-2)}\eta_{\r\s}\eta_{\m\n}+\frac{8}{(d-2)}\frac{1}{p^2}\left(\eta_{\m\n}p_\r p_\s+\eta_{\r\s}p_\m p_\n\right)+}\nonumber\\
  {-}&{\frac{2(\g_1^2-\r_1)}{\g_1^2}\frac{1}{p^2}\left(\eta_{\r\m}p_\s p_\n+\eta_{\r\n}p_\s p_\m+\eta_{\s\m}p_\r p_\n+\eta_{\s\n}p_\r p_\m\right)+}\nonumber\\
  {-}&{\frac{8(2\g_1^2+(d-2)\r_1)}{(d-2)\g_1^2}\frac{1}{p^4} p_\r p_\s p_\m p_\n\Bigg)}.
\end{align}

We are now ready to calculate the 1PI diagram in \cref{fig:1} in dimensional regularization.
After substituting in \cref{diag} for the obtained vertex and propagator, the resulting integral can be expressed as a linear combination of integrals of the type,
\begin{equation}
  \int_{}^{}\frac{d^{d}\,q}{(2 \pi)^d} \frac{1}{q^2(q+p)^2}\frac{P^{(\alpha,\beta)}(q,q+p)}{(q^2)^\alpha((p+q)^2)^\beta},
\end{equation}
where, $\alpha$ and $ \beta $ are non-negative integers such that $0\leq \alpha+\beta\leq4$ and $P^{(\alpha,\beta)}(q,q+p)$ is a polynomial in $q^\mu$ and $(p+q)^\mu$ of dimension $2(\alpha+\beta)$.

The value of such integrals can be consulted in \cite{pascualtarrach} or elsewhere, and setting $d=4+2 \epsilon$ yields,
\begin{align}
  &\qquad\qquad \Gamma_L(p)=-\frac{3 i \left(3 \gamma_1^4+\rho_1^2\right)}{2 \pi ^2 \gamma_1^4 \epsilon}+\nonumber\\
  &+\frac{i \left(9 (1+4 \gamma_E) \gamma_1^4-6 \gamma_1^2 \rho_1+12 \left(3 \gamma_1^4+\rho_1^2\right) \ln(\frac{-p^2}{\mu^2})+(5+12 \gamma_E) \rho_1^2\right)}{8 \pi ^2 \gamma_1^4}+ O\left(\epsilon\right)^{}.\label{37}
\end{align}
{Where, in \cref{37}, we have set $\kappa=1$, and $\gamma_E$ is the Euler-Mascheroni constant.}

Let us now move on and obtain $\Gamma_L(p)$ in \cref{diag} for the BRST formulation of \cite{Kugo}. Now the vertex $\tensor{V}{_L^{\mu_1}^{\nu_1}^{\mu_2}^{\nu_2}}$ reads
\begin{equation}
V_{L}^{\mu_1\nu_1\mu_2\nu_2}={\frac{1}{4}(\eta^{\mu_1\nu_1}\eta^{\mu_2\nu_2}-\eta^{\mu_1\mu_2}\eta^{\nu_1\nu_2}-\eta^{\nu_1\mu_2}\eta^{\mu_1\nu_2})},
\label{Kugovertex}
\end{equation}
for it is $\sqrt{-g(x)}$ {rather} than $g(x)$ which couples to $\lambda(x)$. The graviton propagator of \cite{Kugo} --see (3.15), therein-- is retrieved by setting
\begin{equation*}
  A_1=1,\quad A_2=-1,\quad A_3=2,\quad A_4=-1,\quad A_5=-4,
\end{equation*}
in (\ref{gravitonprop}). For this choice of $A_i$s and the vertex in (\ref{Kugovertex}), $\Gamma_{L}(p)$ turns out to be equal to
\begin{equation}
-\frac{9 i}{32 \pi^2 \epsilon}-\frac{3(11+12 \gamma)
 i}{128 \pi^2}-\frac{9 i}{32 \pi^2}\ln\Big(\frac{p^2}{\mu^2}\Big)+{O}(\epsilon),
\label{generatedcont}
\end{equation}
where $d=4+2\epsilon$ and metric signature is $(-,+,+,+)$. {Again, we} have set $\kappa=1$.

Had we chosen to define the theory \`a la {Wilson} by assuming, for example, that all fields vanish when their momentum is bigger than the UV cutoff scale, $\Lambda_{UV}$, then the integration of the fast graviton modes with momentum
\begin{equation}
p\in[\Lambda, \Lambda_{UV}],
\end{equation}
would give rise for the low energy modes $\l_{low}(x)$ (that is, those whose Fourier transform vanishes when $p>\Lambda$) to
\begin{equation}
c\ln\left({\Lambda_{UV}\over \Lambda}\right) \kappa^2\int d^4 x\,\frac{1}{2}\, \l_{low}(x)^2,
\end{equation}
where $c=\frac{3 \left(3 \gamma_1^4+\rho_1^2\right)}{ \pi ^2 \gamma_1^4}$ and $c=\frac{9}{16 \pi^2}$, respectively, for the two BRST formulations of unimodular gravity we have discussed.
The result that we have obtained above means that the low energy modes $\l_{low}(x)$ do not work as a Lagrange multiplier so that the unimodularity condition is not imposed on the corresponding graviton low energy modes.
\section{A comment on GR in the unimodular gauge.}
The phenomena discussed here also affect GR  {in} the unimodular gauge as discussed in \cite{Baulieu}. The propagator of GR has the same general form discussed above, with
\begin{equation*}
A_1=-1,\,
A_2={2\over d-2},\,
A_3=-{4\over d-2}\nonumber,\,
A_4=1-\a\nonumber,\,
A_5={4(2+\a(d-2))\over d-2}.
\end{equation*}

This leads to the UV divergent result for the corresponding diagram
\begin{equation}
-i{3(3+\a^2)\over 32 \pi^2 \e}+i{9+36\g+\a(-6+\a(5+12\g))\over 128\pi^2}+i{3(3+\a^2)\over 32 \pi^2}\ln\,\Big({p^2\over \m^2}\Big)+O(\e).
\end{equation}
The MS renormalization implies a counterterm
\begin{equation}
{3(3+\a^2)\over 32 \pi^2 \e}\int d^4 x \frac{1}{2}\,\l(x)^2,
\end{equation}
spoiling the working of the {L}agrange multiplier as such.
It is worth pointing out that this counterterm does not jeopardize the BRST quantization of GR, owing to the fact that it is BRST exact.
What it means is that the gauge fixing fermion was not general enough, which is always a delicate issue for theories like GR that are not renormalizable by power counting.

A similar computation can be carried out for GR in the unimodular gauge as defined in \cite{KugoGR}. The result that one obtains is  \cref{generatedcont}, for it   turns out that the graviton propagator of \cite{KugoGR} is the same as the graviton propagator from the unimodular theory of \cite{Kugo}.

\section{A remark on the exponential parametrization.}
Let us present a formal proof of the fact that this third approach of the Introduction does not suffer from this sickness, at least if the unimodularity condition is implemented by adding to the action the term
\begin{equation*}
\int d^d x \,{1\over \kappa}\l(x) h(x).
\end{equation*}
$\lambda(x)$ being a Lagrange multiplier and $h(x)=h^{\mu}_{\mu}(x)$.

Now, let us set $\kappa=1$ and define the partition function as follows
\begin{equation*}
\begin{array}{l}
{Z[J_{\m\n}(x), j(x),\cdots]=e^{i W[J_{\m\n}(x), j(x),\cdots]}=}\\[4pt]
{\int{\cal D} h_{\m\n}{\cal D \l}\cdots\,e^{i S+i\int d^4 x \l(x)h(x)+i(J^{\m\n}(x)h_{\m\n}(x)+j(x)\l(x))\,+\,\cdots},}
\end{array}
\end{equation*}
where $h(x)\equiv \bg^{\m\n} h_{\m\n}{(x)}$ and the dots stand for contributions involving  other fields and external sources but no { $\lambda (x)$}.
An identity can be easily written as
\begin{equation}
0=\int{\cal D} h_{\m\n} {\cal D \l}\cdots\,{\frac{\delta}{\delta \lambda(x)}}\left(e^{i S+i\int d^4 x \l(x)h(x)+i(J^{\m\n}(x)h_{\m\n}(x)+j(x)\l(x))\,+\,\cdots}\right)=i\langle h(x)+j(x)\rangle,
\end{equation}
where as usual
\begin{equation}
\langle F(x)\rangle\equiv \int{\cal D} h_{\m\n} {\cal D \l\cdots}\,F(x)\,e^{i S+i\int d^4 x \l(x)h(x)+i(J^{\m\n}(x)h_{\m\n}(x)+j(x)\l(x)+\cdots)}.
\end{equation}
This Ward identity can be written as
\begin{equation}
-i\bg^{\m\n} {\d Z \over \d J^{\m\n}}+j(x),Z=0=\bg^{\m\n} {\d W \over \d J^{\m\n}}+j(x).
\end{equation}
Define now the 1PI effective action as
\begin{equation}
\Gamma[h_{\m\n},\l,\cdots]\equiv W[J_{\m\n}, j,\cdots]-\int d^4 x \left(J^{\m\n} h_{\m\n}{(x)} + j\, \l{(x)}\right)\,+\,\cdots).
\end{equation}
Then,
\begin{equation}
\bg^{\m\n}h_{\m\n}(x)-{\d \Gamma\over \d \l(x)}=0.
\end{equation}
It follows that the full dependence on the Lagrange multiplier is captured by
\begin{equation}
\Gamma=\tilde{\Gamma}+\int d^4 x \l(x)\bg^{\m\n}h_{\m\n}(x),
\end{equation}
where
\begin{equation}
{\d \tilde{\Gamma}\over \d \l(x)}=0.
\end{equation}
\section{Comments and Conclusions{.}}

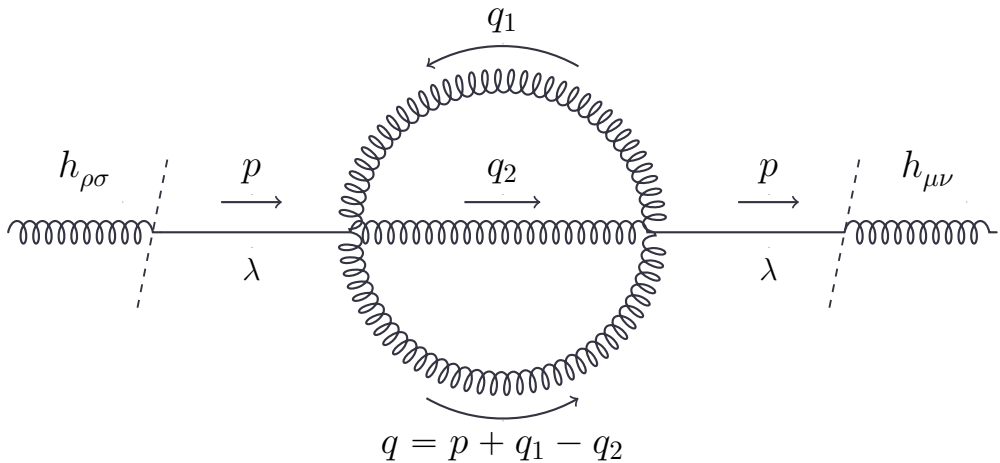
\begin{figure}[h!]
  \centering
  \begin{tikzpicture}[scale=2,
   decoration={coil,amplitude=4.25,segment length=4.75}]
   \filldraw [black] (-2.65,0.25) circle (0pt) node[anchor=south]{\large$p$};
   \draw[thick, ->,blue!7!black!80](-2.85,0.2) -- (-2.45,0.2);
   \filldraw [black] (0.75,0.25) circle (0pt) node[anchor=south]{\large$p$};
   \filldraw [black] (-2.65,-0.1) circle (0pt) node[anchor=north]{$\lambda$};
   \filldraw [black] (0.75,-0.1) circle (0pt) node[anchor=north]{$\lambda$};
   \draw[thick, <-,blue!7!black!80](0.95,0.2) -- (0.55,0.2);
   \draw[thick, ->,blue!7!black!80](-0.5,1.1) arc (60:120:1);
   \draw[thick, <-,blue!7!black!80](-0.5,-1.1) arc (-60:-120:1);
   \draw[thick, ->,blue!7!black!80](-1.25,0.2) -- (-0.75,0.2) ;
   \draw[spring] (-4.25,0) -- (-3.25,0);
   \draw[line width=0.8,blue!7!black!80] (-3.25,0) -- (-2,0);
   \draw[decorate,line width=0.8,blue!7!black!80] (0,0) arc (0:180:1);
   \draw[decorate, line width=0.8,blue!7!black!80] (-2,0) -- (0,0);
   \draw[decorate,line width=0.8,blue!7!black!80] (-2,0) arc (-180:0:1);
   \draw[line width=0.8,blue!7!black!80] (0,0) -- (1.25,0);
   \draw[spring] (1.25,0) -- (2.25,0);
   \filldraw [black] (-3.5,0.2) circle (0pt) node[anchor=south east]{\large$\tensor{h}{_\rho_\sigma} $};
   \filldraw [black] (2.05,0.2) circle (0pt) node[anchor=south east]{\large$\tensor{h}{_\mu_\nu} $};

   \filldraw [black] (-1,0.25) circle (0pt) node[anchor=south]{\large$q_2$};
   \filldraw [black] (-1,1.25) circle (0pt) node[anchor=south]{\large$q_1$};
   \filldraw [black] (-1,-1.25) circle (0pt) node[anchor=north]{\large$q=p+q_1-q_2$};
   \draw[line width=0.7,blue!7!black!80,dashed] (-3.4,-0.5) -- (-3.2,0.5);
   \draw[line width=0.7,blue!7!black!80,dashed] (1.15,-0.5) -- (1.35,0.5);
  \end{tikzpicture}
   \caption{Kinetic contributions to $\lambda(x)$ .}\label{fig:2}
   \end{figure}

Let us {begin} by pointing out that aside from the diagram here studied, one can expect higher-loop diagrams making other terms involving $\lambda(x)$ present in the action, following Gell-Mann's {\em totalitarian principle:} \textit{whatever is not forbidden is compulsory}.
In particular, we expect diagrams such as those in \cref{fig:2} {give rise} to kinetic terms for the $\lambda(x  )$ field: the 1PI diagram in \cref{fig:2} is quadratically divergent by power-counting.

 At any rate, we have shown in this work that the imposition of the unimodular constraint in UG {through} a Lagrange multiplier does not survive, in general, quantum effects; unless the exponential parametrization is used appropriately. These generate, in general, (UV divergent) quadratic (and higher) order terms in the multiplier, as well of kinetic energy contributions, all of which spoil the work of the Lagrange multiplier as such.

The only reasonable hypothetical cancellation of those diagrams would be in an appropriately quantized unimodular supergravity \cite{Anero}.
\par

	\section{Acknowledgements.}
One of us (EA) acknowledges useful e-mail discussions with Luis \'Alvarez-Gaum\'e and Gia Dvali.
	We acknowledge partial financial support by the Spanish MINECO through the Centro de excelencia Severo Ochoa Program under Grant CEX2020-001007-S funded by MCIN/AEI/10.13039/501100011033
	All authors acknowledge the European Union's Horizon 2020 research and innovation programme under the Marie Sklodowska-Curie grant agreement No 860881-HIDDeN and also byGrant PID2019-108892RB-I00 funded by MCIN/AEI/ 10.13039/501100011033 and by ``ERDF A way of making Europe''.
	\newpage
	\appendix
	\newpage

\end{document}